\documentstyle[12pt,epsfig]{article}
\def \Pom {{\hspace{ -0.1em}I\hspace{-0.25em}P}}
\def \Dip {{\hspace{-0.1em}I\hspace{-0.25em}D}}
\begin{document}
\begin{titlepage}
\begin{flushright}
ITP-96-13E
\end{flushright}
\vskip 2.cm
\centerline{\bf PROTON DIFFRACTION DISSOCIATION AND UNITARITY}

\vskip 0.4cm
\centerline{L.L. Jenkovszky, E.S.Martynov and F.Paccanoni}
\vskip 0.3cm

\begin{center}
{\it Bogoliubov Institute for Theoretical Physics,\\
Kiev 143, Ukraine}\\
E-mail: jenk@gluk.apc.org, \, martynov@gluk.apc.org, \, paccanoni@padova.infn.it
\end{center}
\vskip 15.0pt

\begin{abstract}
It is shown that the dipole pomeron model of single diffraction
dissociation -- contrary to the case of the supercritical
pomeron -- is compatible with the inequality
$\sigma^{SD}\leq\sigma^{tot}$, imposed by unitarity, provided the triple
pomeron coupling satisfies certain conditions. With the adopted
approximation the model considered as a parcticular solution
of the triple pomeron decoupling problem. Explicit forms of such a
coupling and a qualitative comparison with the experimental data on
single DD are presented. The modified factorization properties of the
model are also discussed.

\end{abstract}
\end{titlepage}
\newpage
\vskip 2.truecm
\section{Introduction}
~

The renewed interest in hadrons diffraction dissociation (DD),
observed recently has its origin in a different class of events,
namely diffractive deep inelastic scattering, in which the notion
of "pomeron-proton" scattering, or "pomeron flux" is introduced in
terms of the single DD cross section (see e.g. \cite{Kai.-Ter.} -
\cite{Ing.-Schl.} and \cite{Goul.} for a recent presentation of the
problem).

Let us remind that the differential cross section for single DD
in the triple-Regge kinematical limit, $M^2\gg s_0, s/M^2\gg $1
(Fig.1) with account for only one, leading (pomeron) trajectory in each
channel is
$$M^2{d\sigma\over{dt dM^2}}=
{\beta_{h\Pom}(0)\beta^2_{h\Pom}(t)G_{3\Pom}(t)\over{16\pi
}}\bigg(\frac{M^2}{s_{0}}\bigg)^{\alpha(0)-1}\bigg(s/M^2\bigg)^{2\alpha(t)-2}$$
$$=f_{\Pom/p}(x_{\Pom}, t)\sigma_{\Pom p}^{tot}(M^2,t),\eqno (1) $$
where
$$\sigma_{\Pom h}^{tot}(M^2,t)=\beta_{h\Pom}(0)G_{3\Pom}(t)
\bigg(\frac{M^2}{s_{0}}\bigg)^{\alpha(0)-1}$$
is the total cross section of the fictious "pomeron-hadron" scattering,
and
$$f_{\Pom/h}(x_{\Pom},t)=\frac{1}{16\pi}\beta^2_{h\Pom}(t)x_{\Pom}^{1-2\alpha(t)},
\qquad x_{\Pom}=M^{2}/s$$
is the so-called  "pomeron flux", the probability
that the hadron emmits a pomeron.

\parbox{7.cm}
{
\begin{center}
\epsfig{file=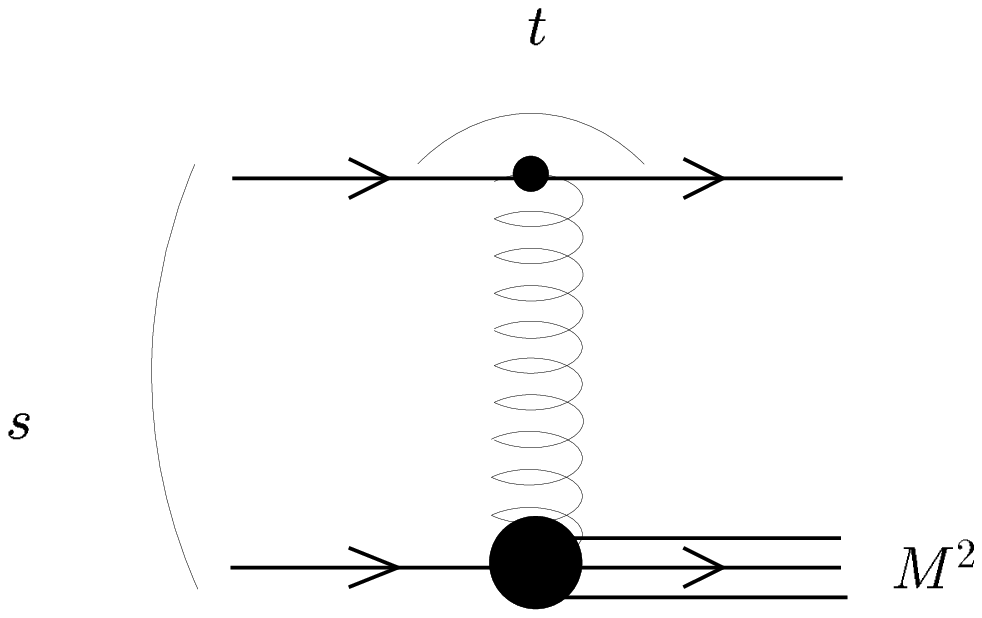,width=6.8cm}
\end{center}
}\ \quad \
\parbox[t]{8.5cm}{

\vskip 1.5cm
Fig.1. Hadron diffraction dis\-so\-cia\-tion.
}

There is an old but still topical problem, namely the
problem of pomeron decoupling (see e.g.\cite{PDT}). If
the pomeron is a simple pole with a linear trajectory of unite
intercept, then $\sigma^{tot}(s)\rightarrow const$ at $s\rightarrow
\infty.$ At the same time, if $G_{3\Pom}\neq 0$ at $t=0$ then it
follows from (1) that $\sigma^{SD}\sim lnlns$ at $s\rightarrow \infty.$
Moreover, the cross-sections of more complicated processes (such as
double diffraction, central diffraction production and so on) rise
with energy even faster than $\sigma^{SD},$ thus violating unitarity.
It was suggested that the pomerons are (in the case of simple
poles) decoupled in a $3\Pom$-vertex at $t=0$ and consequently
$d\sigma/dtdM^{2}$ vanishes when $t\rightarrow 0.$ However
later this was shown (see the review \cite{Gang-Roy}) to
contradict the experimental data.  Thus a serious inconsistency
between the theory and experiment emerged.  We hope
that the problem can be solved in the model under consideration.  Here
for simplicity we consider only  single DD. A complete treatment of
all diffractive contributions will be give elsewhere.

In most of the phenomenological applications a "supercritical"
pomeron i.e. one with the intercept beyond unity, $\alpha(0)=1+
\delta,\ \delta>0$ is used. The value of $\delta$ in reactions in
question (involving on mass shell protons) is about $0.1$,
corresponding to a "soft" pomeron according to the widely used
terminology. This model is attractive for its simplicity:
as shown by A.Donnachie and P.Landshoff \cite{D.-L.} , a single power term mimics, in a
limited energy range the contribution of the pomeron that otherwise
is a very complicated object. Due to the smallness of the parameter
$\delta$, the total cross section with such a pomeron does not violate
in a huge energy range the Froissart bound, following from unitarity.
(Unitarity bounds become more problematic however in the case of
diffraction dissociation to be discussed in this paper.) A single
pomeron term has also the virtue of respecting exactly
factorization, crucial in most of the studies of diffractive deep
inelastic scattering.

Consider now proton single DD, to be denoted in what follows as SD.
Prior to the highest energy Tevatron measurements, data on the
integrated DD cross section were well fitted by a supercritical
pomeron with $\delta\approx 0.1$, as discussed above. Such a behaviour,
however conflicts with the data at higher energies, that lie
well below the relevant extrapolations (see Fig. 2). Moreover, the DD
cross section $\sim s^{2\delta}$ rises faster than the total cross
section $\sim s^\delta$ and overshoots the latter already in the
energy range of the present accelerators.

In a recent paper \cite{Goul.} Goulianos suggested a piece-wise
"unitarization" by introducing a threshold in
energy with different normalizations for the pomeron flux below and
above the threshold. Consequently, the DD cross section gets a "knee"
near that threshold with an abrupt change in the rate of increase.

Although data on DD can be fitted in this way, one can hardly imagine
the dynamics to be discontinuous. Moreover, the inequality
$\sigma^{SD}(\sim s^{2\delta})\leq\sigma^{tot}(\sim s^\delta)$ can not be
satisfied with a supercritical pomeron since integration in $t$
(see below) introduces only logarithmic factors in $t$.
Attempts to solve this problem by summing up an infinite series of unitarity
corrections were undertaking in Refs. \cite{Lev,MaSt}. However
neither the unitarization method nor the final result are yet
conclusive.
Below we show that the inequality
$\sigma^{SD}(\sim s^{2\delta})\leq\sigma^{tot}(\sim s^\delta)$ can be
satisfied for a dipole pomeron. Interestingly, the application of this
bound constrains the form of the triple pomeron vertex. We present a
simple example of such a solution and show also that the model fits the
data.

\medskip
\section{Double $j$-poles in the triple pomeron ($3\Pom$) amplitude}
~

A two-fold, unit intercept pomeron pole is the simplest alternative to
the su\-per\-cri\-ti\-cal pomeron. It provides for rising elastic, inelastic
and total cross sections, as well as the slope
parameter $\sigma^{tot}\sim\sigma^{el}\sim\sigma^{inel}\sim B(s,0)\sim \ln s$
without violating unitarity
bounds (higher multiplicity poles with a linear near $t=0$ trajectories, e.g.
a tripole, are in conflict with
unitarity). Various properties of the dipole pomeron (D\Pom) as well as
their applications to various elastic scattering processes and total
cross sections can be found in \cite{JJDB} and references
therein.

Less explored are the generalizations of the dipole
pomeron to multiparticle reactions.
The first question is: how to write correctly a multiparticle
amplitude with double poles?
To answer it, we first remind that a partial-wave elastic
scattering amplitude with a double pomeron pole can be written in the
$(j,t)$-representation as
$$Im\, a(j,t) =
\frac{\beta^{2}(j,t)}{[j-\alpha_{\Pom}(t)]^{2}}.$$
The amplitude in the
$(s,t)$-representation at large $s$ can be obtained by using a
Mellin transform
$$ A^{\Dip}(s,t) = \frac{1}{2\pi
i}\int\limits_{C-i\infty}^{C+i\infty}dj e^{\xi
(j-1)}\eta(j)\frac{\beta^{2} (j,t)}{[j-\alpha_{\Pom}(t)]^{2}} = $$
$$\frac{1}{2\pi i}\frac{d}{d\alpha_{\Pom}(t)}\int\limits_{C-i\infty}^{C+
i\infty} dj\,e^{\xi(j-1)}\eta(j)\frac{\beta^{2} (j,t)}{j-\alpha_{\Pom}(t)} =
\frac{d}{d\alpha_{\Pom}(t)}A^{\Pom}(s,t) ,$$
where $\xi=\ln (s/s_{0}),\quad s_{0}=const$ and $\eta(j)$ is the signature
factor. The definition of the amplitude $A^{\Pom}(s,t)$ corresponding
to a simple $j$-plane pole  is standard.  For not too large $|t|$
the amplitude $A^{\Dip}(s,t)$ can be written in the form
$$A^{\Dip}(s,t) = i\biggl [G(t)\ln(-is/s_{0}) + \tilde G(t)\biggr
]\bigg (-is/s_{0}\bigg )^{\alpha_{\Pom}(t)-1}, \eqno (2)$$
where we have substituted
$$\eta(j)=-\frac{1+e^{-i\pi j}}{\sin(\pi
j)}=i\frac{1}{\sin(\pi j/2)}(-i)^{j-1},$$
and the $1/sin(\pi j/2)$ factor has been absorbed by $G(t)$
and $\tilde G(t).$
By using the
same procedure, one can write the six-point amplitude with double
Pomeron poles in all $t$-channels (the definition of $t_{i}$ is evident
from Fig.2)

\parbox{8.5cm}{

\vskip -1.cm
\begin{center}
\epsfig{file=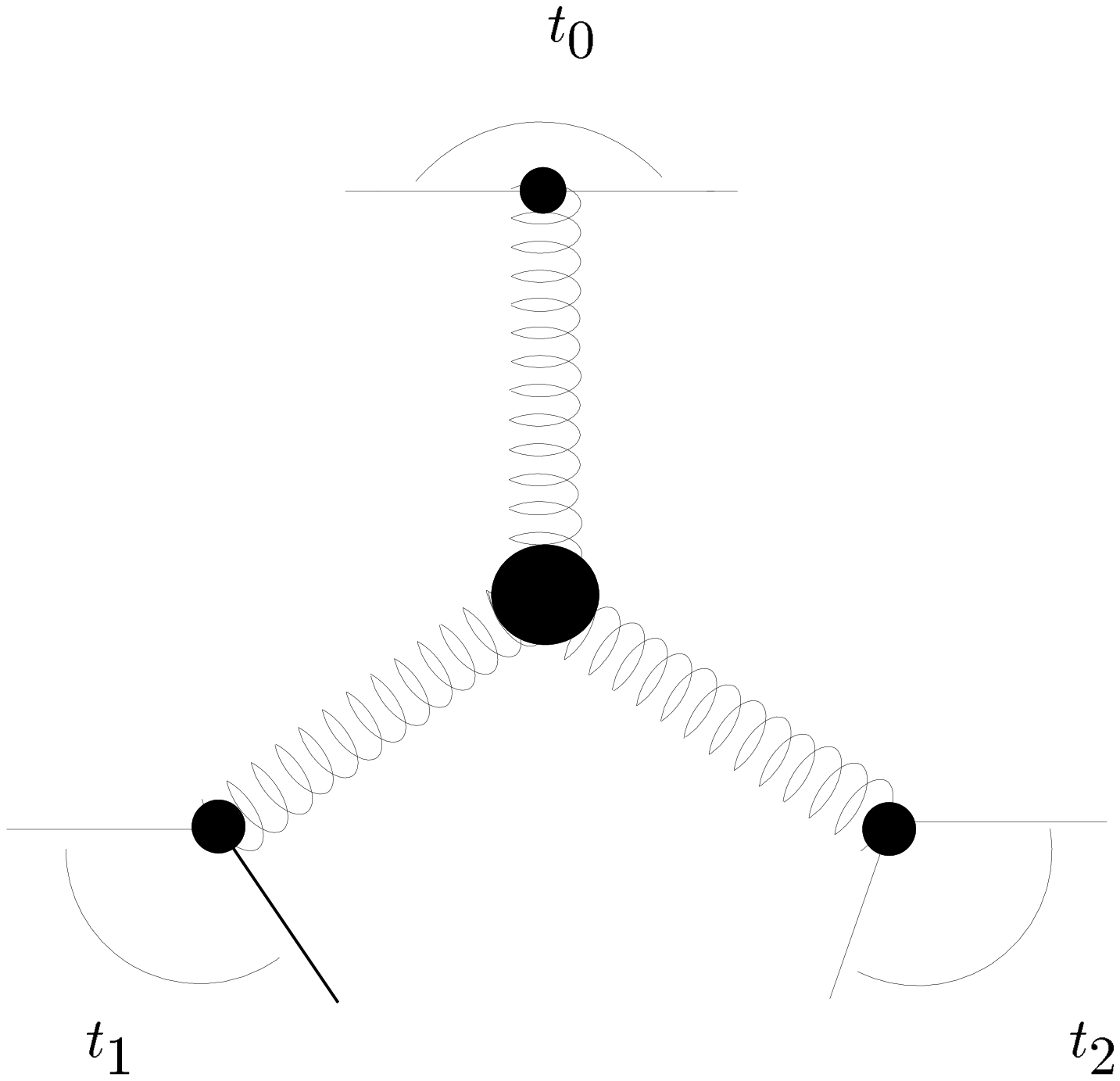,width=8.cm}\ \qquad \
\end{center}}
\parbox[t]{7.7cm}{

\vskip 2.cm
Fig.2. $3\Pom$-diagram: the va\-riab\-les $t_{i}.$}
\vskip -1.cm
$$A^{\Dip}_{6}(s,M^{2},t)
= \frac{d}{d\alpha_{0}} \frac{d}{d\alpha_{1}}\frac{d}{d\alpha_{2}}
A^{\Pom}_{6}(s,M^{2},t_{0},t_{1},t_{2})\mid_{t_{0}=0,\,t_{1}=t_{2}=t;
\,\alpha_{i}=\alpha_{\Pom}(t_{i})}, \eqno (3)$$
where $A^{\Pom}_{6}(s,M^{2},t_{0},t_{1},t_{2})$ is the "contribution" of
a simple pomeron pole in each of the $t$-channels.

Making use of the generalized optical theorem, we obtain for the
single diffractive cross-section (for simplicity we consider
all hadrons identical)
$$M^{2}\frac{d\sigma}{dt dM^{2}} =
\frac{1}{16\pi} \frac{d}{d\alpha_{0}}
\frac{d}{d\alpha_{1}}\frac{d}{d\alpha_{2}}\times
\beta(t_{0},\alpha_{0})\beta(t_{1},\alpha_{1})\beta(t_{2},\alpha_{2})$$
$$G_{3\Pom}(t_{0},t_{1},t_{2};\alpha_{0},\alpha_{1},\alpha_{2})
\biggl(\frac{s}{M^{2}}  \biggr)^{\alpha_{1}+\alpha_{2}-2}
\biggl(\frac{M^{2}}{s_{0}} \biggr)^{\alpha_{0}-1}
\mid_{t_{0}=0,\,t_{1}=t_{2}=t;\,\alpha_{i}=\alpha_{\Pom}(t_{i})}.$$
Here $G_{3\Pom}(t_{0},t_{1},t_{2};\alpha_{0},\alpha_{0},\alpha_{0})$ is a
generalization of the usual triple pomeron, $3\Pom$-vertex. Now we
consider this function in more details.

Evidently, $G_{3\Pom}$ can not be constant, since in this case
the integrated (over $M^{2}$ and $t$)
cross-section $$\sigma^{SD}=
\int\limits_{\xi_{0}}^{\xi-\xi_{0}}d\xi_{1}\int\limits_{-\infty}^{-|t|_{min}}
dt\,\,\frac{d\sigma}{dtd\xi_{1}}\propto \xi^{3}=\ln^{3}(s/s_{0}),$$
would violate the unitary inequality $\sigma^{SD}\leq \sigma^{tot}.$
We remind that in
the dipole Pomeron model $\sigma^{tot}\propto \ln(s/s_{0})$.

Here and in what follows we use the
notation $\xi_{1}=\ln(s/M^{2});\, \xi_{0}$ is a large constant
restricting the region where Regge behaviour is valid. The upper limit
of integration over $t$, ($-|t|_{min}$), generally speaking depends on
$M^{2}/s$, but $|t|_{min}\sim m^2 (M^{2}/s)^{2}\ll 1$ in the region under
consideration ($m$ is the proton mass), so it can be set zero.

Thus, the function $G_{3\Pom}$ should satisfy
the following quite general and natural restrictions:
\begin{enumerate}
\item Symmetry in $t_{1}$
and $t_{2};$
\item $d\sigma/dtdM^{2}\neq 0$ at $t=0;$
\item
Positivity of the cross-section $d\sigma/dtdM^{2}$ at any
$s,t,M^{2};$
\item Unitarity bound $\sigma^{SD}\leq \sigma^{tot}$.
\end{enumerate}

It is easy to see that condition 3) can not be satisfied if $G_{3\Pom}$ is a
linear function of $t_{i},\ \omega_i$ and $t_i$. Below we consider the
case of lowest powers in $t_{i}$ and $\omega_{i}\equiv
\alpha_{\Pom}(t_{i})-1$ compatible with all of the imposed conditions
on $G_{3\Pom}$, namely
$$G_{3\Pom}(t_{0},t_{1},t_{2};\alpha_{0},\alpha_{0},\alpha_{0}) =
G_{3\Pom}^0\exp[\bar b(t_{0}+t_{1}+t_{2})]$$
$$\times [\omega_{0}+g_{1}(\omega_{1}+\omega_{2})+g_{2}\alpha'_{\Pom}(t_{1}+t_{2}]
[\omega_{0}+\tilde g_{1}(\omega_{1}+\omega_{2})+\tilde g_{2}\alpha'_{\Pom}(t_{1}+
t_{2})],\eqno (4)$$
where $\omega_{i}=\alpha_{\Pom}(t_{i})-1$ and $\alpha'_{\Pom}$ is the slope
of the Pomeron trajectory (inserted to make the parameters $g_{2}$
and $\tilde g_{2}$ dimensionless). At large $t_{i},$ the
triple-pomeron vertex $G_{3\Pom}$ may have a complicated
dependence on $t_{i}$, however for the present purposes, here only
its small-$t_{i}$ behaviour will be essential.

One can easily obtain a general expression for the differential
cross-section, corresponding to Exp.(4) for $G_{3\Pom}$
$$\frac{d\sigma}{dtd\xi_{1}} = \frac{1}{16\pi}\beta^{3}(0)G_{3\Pom}
\exp[(B+2\alpha'_{\Pom}\xi_{1})t]$$
$$\times \bigl \{G_{1}(\xi-\xi_{1})\xi_{1}^{2}(2\alpha'_{\Pom}t)^{2}+
G_{2}(\xi-\xi_{1})\xi_{1}(2\alpha'_{\Pom}t)+
G_{3}\xi_{1}^{2}(2\alpha'_{\Pom}t)+
G_{4}\xi_{1}+G_{5}(\xi-\xi_{1})\bigr
\},\eqno (5)$$
where
$$G_{1}=(g_{1}+g_{2})(\tilde g_{1}+\tilde g_{2}),\quad G_{2}=2[g_{1}(\tilde g_{1}+\tilde g_{2})
+\tilde g_{1}(g_{1}+g_{2})],$$
$$G_{3}=(g_{1}+g_{2}+\tilde g_{1}+\tilde g_{2}),\quad G_{4}=2(g_{1}+\tilde g_{1}),\quad
G_{5}=2g_{1}\tilde g_{1}.$$
$$B=2b+2\bar b\qquad
\mbox{if } \qquad
\beta(t)=\beta(0)e^{bt}.$$

After  integration in $t$ and $\xi_{1}$ we find that the term
violating the inequality $\sigma^{SD}\leq \sigma^{tot}$ (it behaves
like $\xi\ln\xi$ for $\xi\rightarrow \infty$) has a factor
$2G_{1}-G_{2}+G_{5}$. Hence, by setting
$$2G_{1}-G_{2}+G_{5} = 0,\eqno (6)$$
we obtain
$$\sigma^{SD} = C_{1}\ln(s/s_{0}) + C_{2}\ln(\ln(s/s_{0})) + C_{3}
+ \cdots \eqno (7)$$

We did not specify the constants $C_{i}$ (one can easily express
them through $g_{i},\tilde g_{i}$) because a more general form for the
amplitude can be constructed by taking into account the combination of
simple and double pomeron pole.  In particular, one can use
$$\frac{d}{d\alpha_{i}}\quad \rightarrow \quad \phi_{1}(t_{i})+\phi_{2}(t_{i})
\frac{d}{d\alpha_{i}}$$ instead of the simple derivative used here with
the relevant modification of $\sigma^{SD}.$

To illustrate the applicability of the model and anticipating future
detailed fits to the data, here we present only a simple fit to
$\sigma^{SD}$ by an approximate choice of the values of
the free parameters, without applying the minimization procedure.
We give two examples. The first one corresponds to the asymptotic
expression (7) with $C_{1}=0.06mb,\quad C_{2}=3mb,\quad C_{3}=0.$
The second curve comes from the expression
$$\sigma^{SD} = C_{1}lns + C_{2}ln(lns+B) + C_{3} +
\frac{C_{4}}{lns+B},\eqno(8)$$
$$C_{1}=0.2mb,\quad C_{2}=2.9mb,\quad C_{3}=-1.6mb,\quad
C_{4}=-12mb,\quad B=6 $$
which takes into account the
preasymptotic terms as well. In our opinion, a detailed comparison
with the
data on the cross-section $d\sigma^{SD}/dtdM^{2}$ should
be made because the model has a quite complicated form of the
"$3\Pom$-vertex".

\vskip 1.5cm
\parbox{7.5cm}
{\hskip -.5cm
\epsfig{file=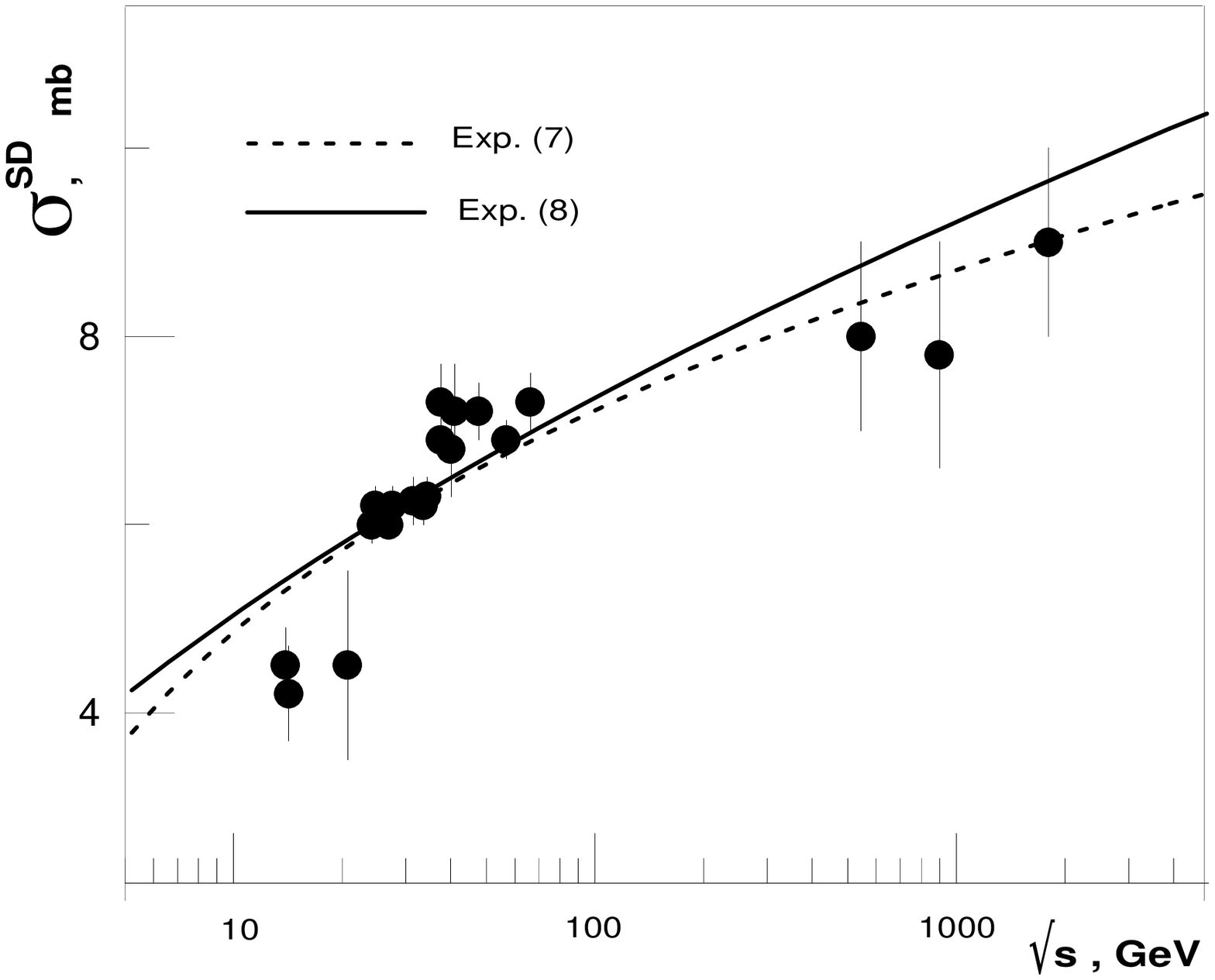,width=7.cm}}\ \qquad \
\parbox{7.5cm}{
Fig.3. Semi-qualitative fit to the data in the dipole pomeron model
presented in this paper. The values of the adjustable parameters
were chosen without any minimization procedure - just to illustrate
the idea.}

\section{Factorization}
~

Strictly speaking, factorization -- in its simplest form -- is
satisfied only if the relevant exchange (pomeron) is a monom
, as in the case of the D-L model \cite{D.-L.}, in which the
scattering amplitude $\sim s^\delta$. Actually, this simplicity is
never realized in the nature, since apart form the rising part,
diffraction (=pomeron exchange) unavoidably contains also a
constant component, required e.g. by
duality and/or experimental data at the present, not asymptotical energies.

Factorization is efficient as a crude approximation to reality, but
in any quantitative analysis of the available data one
should account for the departu\-re from factorization. The problems is
technical, rather than conceptual, but it may become crucial for
a correct analysis of the data.

The dipole pomeron model always implies the presence of at least two
terms in the amplitude, one corresponding to a simple pole
exchange (constant cross sections), the other one -- to a dipole
(logarithmically rising term). To illustrate the modified
factorization form, let us write a simplified example of the elastic
scattering
$$A^{\Dip}(s,t) = \frac{d}{d\alpha}\beta(t)\Pom (s,t)\beta(t) = $$
$$\beta(t)\Dip (s,t)\beta(t) + \beta'(t)\Pom (s,t)\beta(t) +
\beta(t)\Pom (s,t)\beta'(t),$$
where
$$\Dip (s,t)=\ln(-is/s_{0})\Pom (s,t),\qquad
\Pom (s,t)=\big(-is/s_{0}\big)^{\alpha_{\Pom}(t)-1}.$$

Exact factorization is restored at "asymptotic" energies, when the
second and third terms (simple pole contributions) can be neglected. When
does it happens -- depends on the actual values of the fitted
parameters.

The same problem -- but in a more complicated form -- appears in the
amplitude of the DD in a dipole pomeron model.
The "generating" amplitude $A_{6}^{\Pom}$ has the factorized form but
the genuine amplitude $A_{6}^{\Dip}$ is represented by a sum of a
few terms and does not factorized. Nevertheless factorization is
restored at a far asymptotics and at $t\neq 0$, when the leading
term with $(\xi-\xi_{1})\xi_{1}^{2}$\, (see Exp.(4)) dominates.

An important conclusion  following from the above arguments is that
the "pomeron flux" can be defined only in the asymptotic sense. It
may be that the factorized form (1)  of the pure hadronic
cross-section as well as an analogous form for diffractive DIS are only
approximate at available energies. So we must be careful about
the conclusions on the quark-gluon content of the pomeron
relying too much on the factorization and the concept of the pomeron
flux.

The diffractive structure function $F_{2}^{4(D)}$ is usually calculated
making use of the Pomeron structure function $F_{2}^{\Pom}$, namely
$$\frac{d\sigma^{DDIS}}{dtdx_{\Pom}dxdQ^{2}} =
\frac{4\pi\alpha^{2}}{xQ^{4}}(1-y+\frac{y^{2}}{2})F_{2}^{4(D)},\eqno(9) $$
$$F_{2}^{4(D)} = F_{2}^{\Pom}(x,Q^{2},x_{\Pom},t)f_{\Pom/p}(x_{\Pom},
t),\eqno(10)$$
where the approximation $R=\sigma_{L}/\sigma_{T}=0$ is implied
for simplicity.  However as we discussed above, the "pomeron
flux" makes no sense in the non-asymptotic region.  How can
$F_{2}^{4(D)}$  be calculated in
this case? We propose a method, based on the idea shown in
Fig.4, and briefly described below.

\begin{center}
\epsfig{file=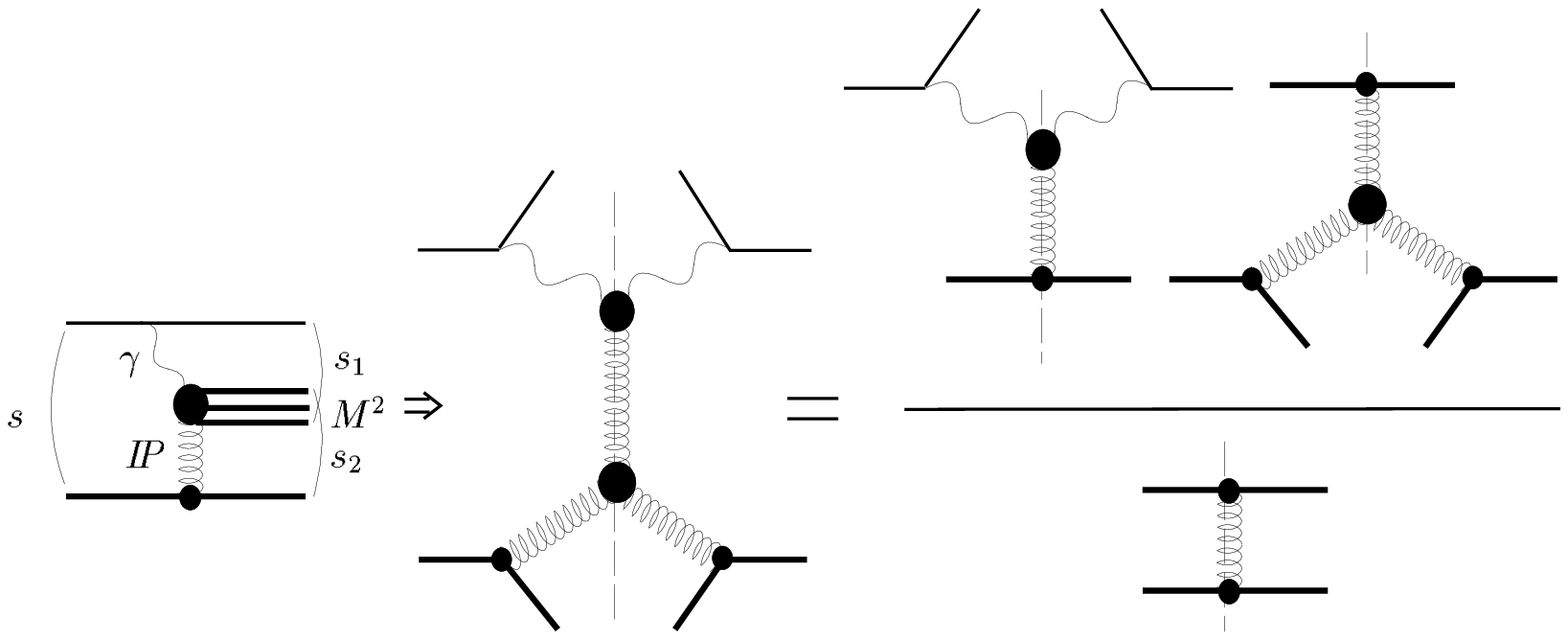,width=16.cm}
\vskip 4.cm

Fig.4. Illustration of the factorization
idea for a simple pole
\end{center}

 Let us consider first the contribution of a single
factorizable simple pole in the
crossed channel to the undetected particles.  In accord with Fig.4 one
may write

$$\frac{d\sigma^{DDIS}}{dtdx_{\Pom}dxdQ^{2}} =
\frac{d\sigma^{DIS}/dxdQ^{2}\cdot
d\sigma^{SD}/dtd\xi^{'}}{\sigma^{tot}_{pp}(s)}\frac{Q^{2}}{Q^{2}(1-x)+m^{2}x},$$
where $\xi^{'}=ln(s/s_{1})$ and $s_{1}$ is defined in Fig.4.
Making use of the well known relations between the
cross-sections and structure functions (Exp.(9) for $F_{2}^{4(D)}$
and its analog for $F_{2}^{p}$) we can express $F_{2}^{4(D)}$ through
the usual nucleon structure function and pure hadronic cross-sections:
$$F_{2}^{4(D)}\propto \frac{1}{\sigma^{tot}_{pp}(s)}F^{p}_{2}(x,Q^{2})
\frac{d\sigma^{SD}}{dtd\xi^{'}}.$$

With account for more  contributions (e.g. double pole,
simple pole, nonleading reggeons etc.) the cross-section and structure
function will be rewritten in the following form
$$\frac{d\sigma^{DDIS}}{dtdx_{\Pom}dxdQ^{2}} =
\frac{Q^{2}}{Q^{2}(1-x)+m^{2}x}\sum\limits_{i}
\frac{d\sigma^{DIS}_{i}/dxdQ^{2}\cdot
d\sigma^{SD}_{i}/dtd\xi^{'}}{\sigma^{tot}_{i}(s)}
,$$
$$F_{2}^{4(D)}\propto
\sum\limits_{i}\frac{1}{\sigma^{tot}_{i}(s)}F^{p}_{2,i}(x,Q^{2})
\frac{d\sigma^{SD}_{i}}{dtd\xi^{'}},$$
where each of the factors in the sums is the partial
contribution to the corresponding quantity
 $$\sigma^{tot} = \sum\limits_{i} \sigma^{tot}_{i}, \quad
F^{p}_{2} = \sum\limits_{i}F^{p}_{2,i}, \quad
\frac{d\sigma^{SD}}{dtd\xi^{'}} = \sum\limits_{i}
\frac{d\sigma^{SD}_{i}}{dtd\xi^{'}}.$$

\section{Conclusions}

In this paper we have presented a model for the pomeron (dipole
pomeron) compatible with unitarity and the experimental data. We have
treated only the simplest case of single diffraction, but results are
encouraging and we believe that the inclusion of more complicated
diagrams, like double DD will complete this study and resolve the
puzzle of the so-called decoupling theorems.

Breakdown of factorization is an essential consequence of this
approach. Let us stress that
the breakdown (or restoration) of factorization in the present model
depends on a non-trivial interplay of the $s-$ and $t-$ dependence.
We remind that
factorizability of the pomeron contribution was a crucial point in
most of the calculation and measurements of diffractive deep
inelastic scattering (DIS). Moreover, in the Ingelmann-Schlein model
for diffractive DIS \cite{Ing.-Schl.},
factorization is an unavoidable ingredient to make the definition of
the pomeron structure function, multiplied by the pomeron
flux (Exps. (9) and (10)), sensible. The introduction of a realistic pomeron
model, unavoidably will modify the simple
factorization of the amplitudes and cross-sections.
Finally, we note the second important source of non-factorizability,
coming from the contribution of secondary (e.g., $f$) trajectories.
Their effect will be treated elsewhere. To conclude, let us notice
that nonfactorizability is observed also experimentally in
diffractive DIS \cite{Rom} and has become a subject of intensive
exploration -- both theoretical and experimental.


 \end{document}